\documentclass[reprint,amsmath,amssymb,aps,prl,showpacs]{revtex4-1}

\usepackage{graphicx}
\usepackage{dcolumn}
\usepackage{bm}
\usepackage{esint}
\usepackage{braket}
\usepackage{multirow}
\usepackage{lineno}
\usepackage{color}

\newcommand{\Yb}{\ensuremath{^{171}\mathrm{Yb}^+~}}
\newcommand{\avg}[1]{\ensuremath{\left\langle#1\right\rangle}}

\begin{document}

\preprint{APS/123-QED}

\title{Quantum Implementation of Unitary Coupled Cluster for\\ Simulating Molecular Electronic Structure
}

\author{Yangchao Shen$^{1}$, Xiang Zhang$^{1}$, Shuaining Zhang$^{1}$, Jing-Ning Zhang$^{1}$, Man-Hong Yung$^{2,1*}$ , Kihwan Kim$^{1*}$}

\affiliation{
$^{1}$Center for Quantum Information, Institute for Interdisciplinary Information Sciences, Tsinghua University, Beijing 100084, P. R. China.\\
$^{2}$Department of Physics, South University of Science and Technology of China, Shenzhen 518055, P. R. China
}

\begin{abstract}
In classical computational chemistry, the coupled-cluster ansatz is one of the most commonly used $ab~initio$ methods, which is critically limited by its non-unitary nature. The unitary modification as an ideal solution to the problem is, however, extremely inefficient in classical conventional computation. Here, we provide the first experimental evidence that indeed the unitary version of the coupled cluster ansatz can be reliably performed in physical quantum system, a trapped ion system. We perform a simulation on the electronic structure of a molecular ion (HeH$^+$), where the ground-state energy surface curve is probed, energies of excited-states are studied and the bond-dissociation is simulated non-perturbatively. Our simulation takes advantages from quantum computation to overcome the intrinsic limitations in classical computation and our experimental results indicate that the method is promising for preparing molecular ground-states for quantum simulation.
\end{abstract}

\maketitle

The central problem in quantum chemistry and molecular physics is to determine the electronic structure and the ground-state energy of atoms and molecules by solving the quantum many-body equations, which is generally intractable due to the exponential scaling to the size of the system. Quantum simulation \cite{Aspuru-Guzik05,Mueck2015,KassalSimulating,Yung2012c,Aspuru-Guzik12,Yung2014} can provide the solution for such ``exponential catastrophe'' problem. The key ingredient of quantum molecular simulation consists of ($i$) ground (excited) -state preparation and ($ii$) energy estimation of the corresponding state \cite{KassalSimulating,Yung2012c}. Recently, the assessed costs for the energy estimation for a well-prepared ground-state in quantum computation have been immensely reduced \cite{Wecker2014,Hastings2015,Poulin2015,McClean2014p,Babbush2015}, indicating that chemistry simulation can be one of the main applications of a quantum computer in near future. However, it is still remaining major obstacle to efficiently and reliably find the molecular ground state, which belongs to the class of extremely hard problems called Quantum Merlin Arthur, the quantum analog of NP-hard problem \cite{Aaronson2009,Whitfield2012}. Recently various theoretical schemes for the ground-state problem have been proposed and proof-of principle experimental demonstrations have been performed including the adiabatic \cite{Farhi01,Kim10,Johnson11} and algorithmic preparations \cite{Laflamme05,Barreiro11,Wineland13,Guo14}.

For the ground-state problem, the developments of conventional quantum chemistry can be adopted to quantum computation. In computational chemistry, it has been the main focus to circumvent the problem by approximating the many-body Schr{\"o}dinger equation and a series of theoretical and numerical methods have been developed. The coupled-cluster method is one of the most prominent $ab~initio$ methods for finding a molecular ground state and it is considered to be the current gold standard \cite{Bartlett2007,shavitt2009many,Pittner2010,atkins2011molecular}. However, the coupled-cluster ansatz is built with non-unitary operation, which leads to drawbacks such as lacking a variational bound on the ground-state energy \cite{shavitt2009many,Pittner2010,atkins2011molecular,Chan2004,Taube2006}. The unitary version of the coupled-cluster methods would perfectly resolve the problem, whereas it is classically inefficient without proper truncation of the infinite series expansion. It has been a long-standing challenge to build an efficient computational scheme for the unitary coupled-cluster (UCC) ansatz. The authors of Refs. \cite{Yung2014, Peruzzo2013} pointed that the UCC ansatz can be efficiently implemented in a quantum computer. In other words, the quantum implementation of the UCC method can outperform the classical computation for the problem of finding molecular ground-state.

\begin{figure}[ht]
\includegraphics[width=0.5\textwidth]{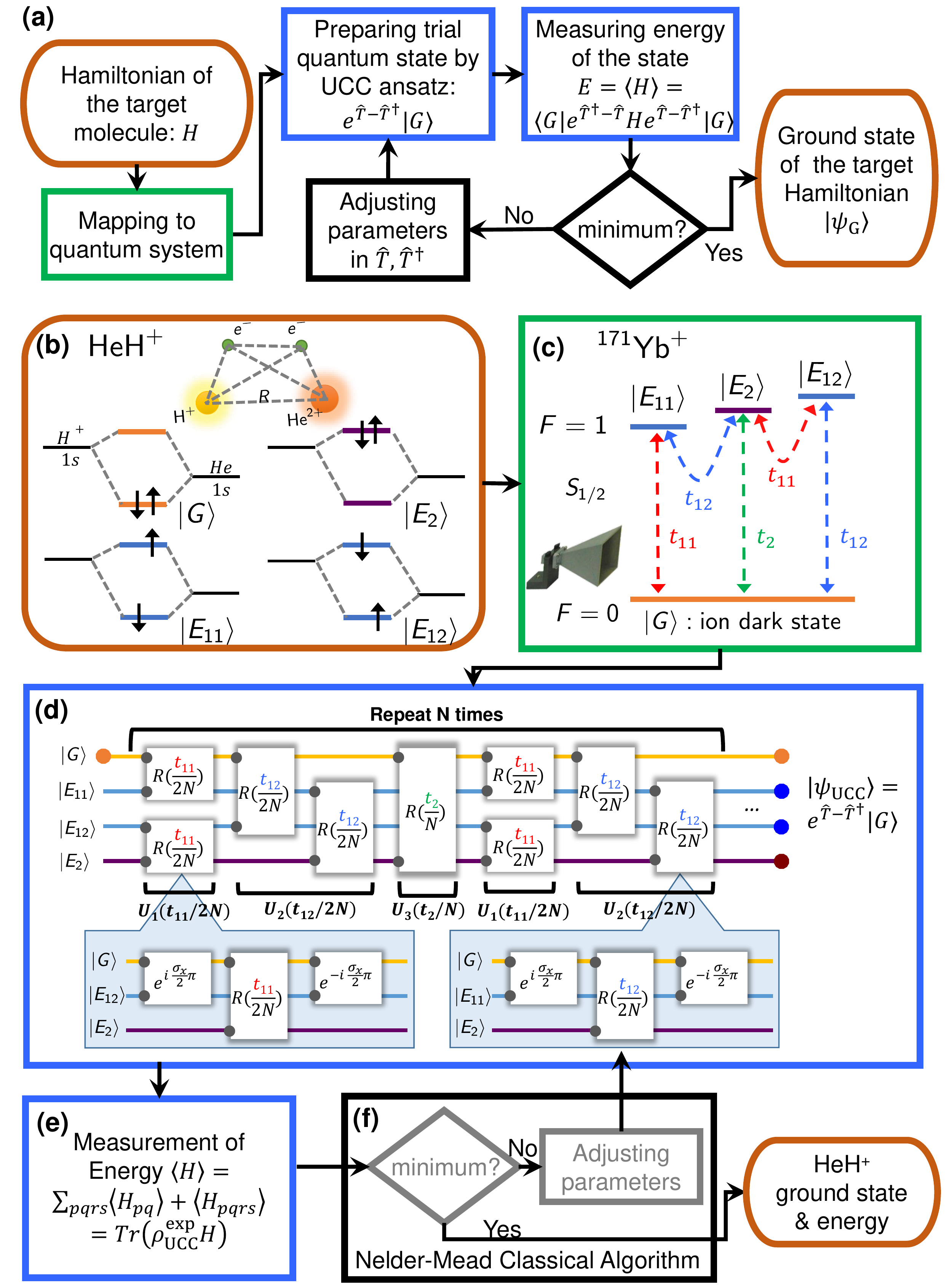}
\caption{\label{fig:Fig1} (a) The conceptual procedure and (b-f) the experimental realization of the UCC simulation to find the ground state of target molecule. (b) The Hartree-Fock basis states of the target molecule, HeH$^+$. (c) The mapping of the basis states on the energy levels of $^{171}$Yb$^+$ including cluster amplitudes $t_{11},t_{12}$, and $t_2$, which are controlled by the duration of microwave pulses. (d) The microwave pulse sequence for the preparation of the UCC ansatz. The effective time evolution operator $e^{T-T^\dag}$ is expanded by the Suzuki-Trotter scheme (see SM. C: Quantum Implementation of the UCC Ansatz). (e) The measurement of energy $\avg{H}$ of the UCC ansatz given cluster amplitudes. (f) A classical minimum search algorithm (Nelder-Mead) is applied to determine the cluster amplitudes for the next UCC ansatz and the final ground state.}
\end{figure}

Recently a variational method of approximating molecular ground states has been experimentally tried in a photonic system \cite{Peruzzo2013}. Due to challenges in the system, however, the variational ansatz employed in the experimental demonstration was not the UCC ansatz, but called 'device ansatz' which is device specific method, therefore is not scalable. As discussed in Ref. \cite{Yung2014}, the UCC ansatz provides a generic and scalable scheme for generating a parameterized state for the variational method and can be implemented efficiently with quantum devices including trapped ions.

Here we report the first experimental realization of the UCC ansatz with a minimal basis, based on quantum simulation in a multi-level of a trapped $^{171}$Yb$^+$ ion. We simulate the electronic structure of a molecular ion (HeH$^+$) \cite{Peruzzo2013,Wang2015} and reliably find the molecular ground-state as well as the corresponding energy by the UCC ansatz and the variational method, which can be considered as an alternative method for the energy estimation \cite{Lanyon10,Du10}. Moreover, we apply the quantum UCC method to compute excited states and chemical-bond softening non-perturbatively.

The coupled-cluster approach is based on the variational method with the trial state of the exponential ansatz in the form of ${e^{ T}}\ket{G}$. Here $\ket{G}$ is a reference state, such as the Hartree-Fock ground state and the cluster operator $T$ is constructed by a sum of $n$-electron excitation operator $T_n$ with the transition amplitudes as variational parameters (see Eq.~(S1) in Supplemental Material (SM)). However, the operator $T$ is not necessarily Hermitian and energy estimation by the ansatz is not guaranteed to bounded by variational theorem \cite{Chan2004,Taube2006}.

The UCC scheme based on the form of the following ansatz
\begin{equation}\label{eq1}
\ket{\psi_{\rm UCC}}=e^{T-T^{\dagger}}\ket{G}
\end{equation}
which apparently provides a solution of the non-Hermitianity problem in the coupled-cluster theory. However, classical implementation of the UCC have intrinsic limitation, e.g., infinite series of the expansion \cite{Taube2006} (see also SM. B: Classical Implementation of UCC Ansatz). As a result, all classical applications of UCC involve some type of truncations with potentially uncontrollable errors. On the other hand, the unitary operator, $U \equiv e^{T - {T^\dag }}$, can be considered as a time-evolution operator, $i.e.$, $U \equiv e^{ - i{H_{\rm eff}}}$, driven by an effective Hamiltonian ${H_{\rm eff}} \equiv i ( {T - {T^\dag }} )$ with a dimensionless time interval set to be $1$. Since the time-evolution is efficiently simulated in a quantum system \cite{Lloyd96}, the quantum implementation of the UCC ansatz can reduce the computational cost much less than the classical requirement.

The whole procedure of finding the ground state of a target molecule is shown in Fig. \ref{fig:Fig1}(a), which is also discussed in Refs. \cite{Yung2014, Peruzzo2013}. After efficiently preparing a trial state with UCC ansatz in a quantum system that maps the classical basis set of the target molecule, we measure the average energy of the state. The preparation of the UCC ansatz and the energy measurement are performed in the quantum system. Based on a classical feedback algorithm, we adjust the parameters, $i.e.$, the cluster amplitudes of the UCC ansatz. We repeat the quantum process of preparation and measurement until we find the variational minimum of the target Hamiltonian.

{\bf Target Molecule} (see also SM. C: The Hamiltonian of HeH$^+$). We choose the helium hydride cation (HeH$^+$) \cite{Peruzzo2013,Wang2015} for the computation of the energy curve. In the second quantization representation with the minimal STO-3G basis-set \cite{atkins2011molecular} from the 1s orbitals of Hydrogen and Helium, the Hamiltonian of HeH$^+$ is described by
\begin{equation}\label{eq3}
H(R)=\sum_{pq}h_{pq}(R)\hat{a}_p^\dagger\hat{a}_q + \frac{1}{2} \sum_{pqrs}h_{pqrs}(R)\hat{a}_p^\dagger \hat{a}_q^\dagger \hat{a}_r \hat{a}_s
\end{equation}
where $R$ is the nuclear separation between hydrogen and helium, $h_{pq}(R)$ and $h_{pqrs}(R)$ are related to one electron and two electron transitions, respectively and the index $p,q,r,s$ stands for the four possible states in our Hilbert space. The terms of $h_{pq}(R)$ and $h_{pqrs}(R)$ are computed numerically by the Hartree-Fock method with the scaling of $O(M^4)$, where $M$ is the number of molecular orbits. The creation and annihilation operators in the Hamiltonian (\ref{eq3}) are mapped to spin Pauli operators by performing the Jordan-Wigner transformation and pairs of Pauli operators are mapped to four-level systems. 
After the mapping, the Hartree-Fock basis for HeH$^+$ consists of the following set of four states, $\{\ket{G}, \ket{E_{11}}, \ket{E_{12}}, \ket{E_{2}}\}$ as shown in Fig.~\ref{fig:Fig1}(b). 

{\bf Mapping of HeH$^+$ on \Yb ion.} The electron excitation operators, which excite the electrons out of the Hartree-Fock ground state, up to two electron excitations are given by,
\begin{equation}\label{eq4}
  {T_1} = {t_{11}}a_{2 \downarrow }^\dag {a_{1 \downarrow }} + {t_{12}}a_{2 \uparrow }^\dag {a_{1 \uparrow }} ~,~ {T_2} = {t_2}a_{2 \downarrow }^\dag a_{2 \uparrow }^\dag {a_{1 \uparrow }}{a_{1 \downarrow }}
\end{equation}
Note that all the terms are spin preserving, and the $t_{11},t_{12},t_{2}$ are in general complex numbers to be determined by an optimization process. After the same mapping process of the Hamiltonian (\ref{eq3}), the effective Hamiltonian ${H_{\rm eff}} \equiv i ( {T - {T^\dag }} )$ for the cluster operators is written as
\begin{eqnarray}\label{eq5}
H_{ \rm eff} &=& i t_{11} \left( \ket{E_{11}}\bra{G} + \ket{E_{2}}\bra{E_{12}} \right) \nonumber \\
&+& i t_{12} \left( \ket{E_{12}}\bra{G} + \ket{E_{2}}\bra{E_{11}} \right)  \\
&+& i t_{2} \ket{E_{2}}\bra{G} +  \textrm{h.c.}. \nonumber
\end{eqnarray}
We realize the effective Hamiltonian $H_{\rm eff}$ in a quantum system of multiple energy levels in trapped \Yb ion. As shown in Fig.~\ref{fig:Fig1}(c), four energy levels in the ground-state manifold of $^2S_{1/2}$ of the $^{171}$Yb$^+$ are employed \cite{Xiang13,xiang15} to map the basis state as $\ket{F=0,m_F=0}\equiv\ket{G}$ and $\ket{F=1,m_F=-1,1,0}\equiv \left\{\ket{E_{11}},\ket{E{_{12}}},\ket{E_{2}}\right\} $, which are separated by $\omega_{\rm HF}-\omega_{\rm z}~,~\omega_{\rm HF}+\omega_{\rm z}$ and $\omega_{\rm HF}$, where the hyper-fine splitting of $\omega_{\rm HF}=\left(2\pi\right)12.642821{\rm GHz}$, Zeeman splitting of $\omega_{\rm z}=\left(2\pi\right)13.586{\rm MHz}$ with the static magnetic field of $B= 9.694 {\rm G}$. 

{\bf Preparation of the UCC Ansatz} (see also SM. D: Quantum Implementation of the UCC Ansatz). The unitary operator $U \equiv e^{ - i{H_{\rm eff}}}$ is implemented as a time evolution of the system with the effective Hamiltonian ${H_{\rm eff}}$ as shown in Fig.~\ref{fig:Fig1}(d). The initialization of the state to $\ket{G}$ is performed by the standard optical pumping technique. The transitions $\left\{ \ket{G} \leftrightarrow \ket{E_{11}}, \ket{G} \leftrightarrow \ket{E_{12}} ,\ket{G}  \leftrightarrow \ket{E_{2}}  \right\}$ are implemented by applying resonant microwaves. The other transitions $\left\{ \ket{E_{11}} \leftrightarrow \ket{E_{2}}, \ket{E_{12}} \leftrightarrow \ket{E_{2}} \right\}$ are achieved by applying composite pulse sequences shown in the insets of Fig.~\ref{fig:Fig1}(d). Consequently, the experimental implementation of the unitary operator $U$ is achieved by the sequence depicted in Fig.~\ref{fig:Fig1}(d), which results from the second-order Suzuki-Trotter expansion. In the experiment, the UCC amplitudes $t_{11},~t_{12},$ and $t_{2}$ are controlled by the durations of the corresponding microwave transitions. We note that for HeH$^{+}$, the amplitudes near the molecular ground state are much smaller than 1 and the errors from small Trotter expansions ($N=2$ in our experiment) are negligible.

\begin{figure}[ht]
\includegraphics[width=0.5\textwidth]{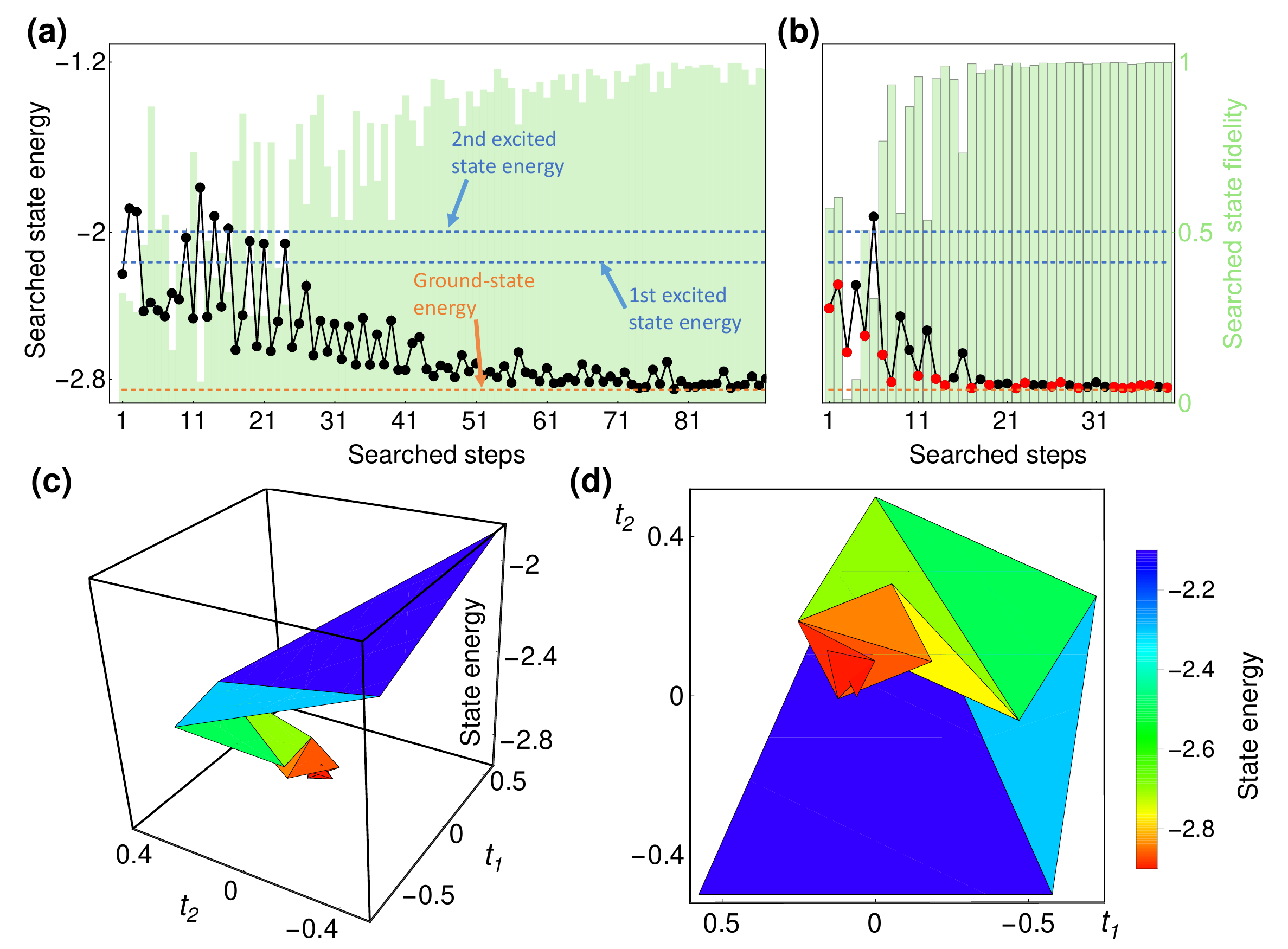}
\caption{\label{fig:Fig2} The search process of the minimum energy at $R = 1.7 a.u.$ assisted by the classical Nelder-Mead algorithm with UCC ansatz. The measured energy $\avg{H}$ (dots) and the fidelity of the prepared state (bars) to the ideal ground state depending on the number of iterations (a) with full six parameters and (b) with two parameters. For both cases, the algorithm converges to the ground-state energy obtained by the exact diagonalization with decent fidelity of the state. Red dots show the successful steps that contribute to the convergence. (c) The side view and (d) the bottom view of the searching process with two parameters for the successful steps. The atomic unit ($a.u.$) is used for all the figures.}
\end{figure}

{\bf Energy Measurement of the UCC Ansatz.} We can obtain the energy $\avg{H}=\sum_{pq} \avg{H_{pq}} + \sum_{pqrs} \avg{H_{pqrs}}$, where $H_{pq}=h_{pq} \hat{a}_p^\dagger\hat{a}_q$ and $H_{pqrs}=h_{pqrs} \hat{a}_p^\dagger \hat{a}_q^\dagger \hat{a}_r \hat{a}_s$, by term-by-term measurements and addition of all of them in the target Hamiltonian (\ref{eq3}). For our case of HeH$^+$ in the minimal basis, it requires 24 times of measurements (SM. C. The Hamiltonian of HeH$^+$), which necessarily needs the information of all the components in the density matrix of the UCC ansatz (see also E. Measurement of the Energy for HeH$^+$). Note that as the system size increases, we do not need the full knowledge of the density matrix of the state for the energy measurement, since the number of terms in the Hamiltonian (\ref{eq3}) scales polynomially \cite{Yung2014,Peruzzo2013}. Since we need the full knowledge of the density matrix for our small scale simulation, we reconstruct the full density matrix $\rho_{\rm UCC}^{\rm exp}$ by the standard quantum state tomography, which requires 15 times of measurement, and obtain the energy by $Tr \left(\rho_{\rm UCC}^{\rm exp} H\right)$. For the relevant components of the density matrix, we repeat the standard measurements up to 1000 times, which give $3.2\%$ projection uncertainty of standard deviation. 

{\bf Classical Minimization Algorithm} The preparation and the measurement of an UCC ansatz are performed in quantum system and the minimization process is supported by classical algorithm. The measured value of $\avg{H}$ for the prepared UCC ansatz is taken as an input for a classical optimization algorithm, which compares it to the previous values and suggests a new set of $\{t_{11},t_{12},t_2\}$ so that the same procedure is repeated until the resulting $\avg{H}$ converges to some value. As a result, we obtain an optimized state with minimal energy for approximating the ground state of HeH$^+$ in the form of an UCC ansatz in Eq.~(\ref{eq1}). In our realization, we use a popular Nelder-Mead minimum search algorithm\cite{thompson2009simulation}. 

\begin{figure}[ht]
\includegraphics[width=0.5\textwidth]{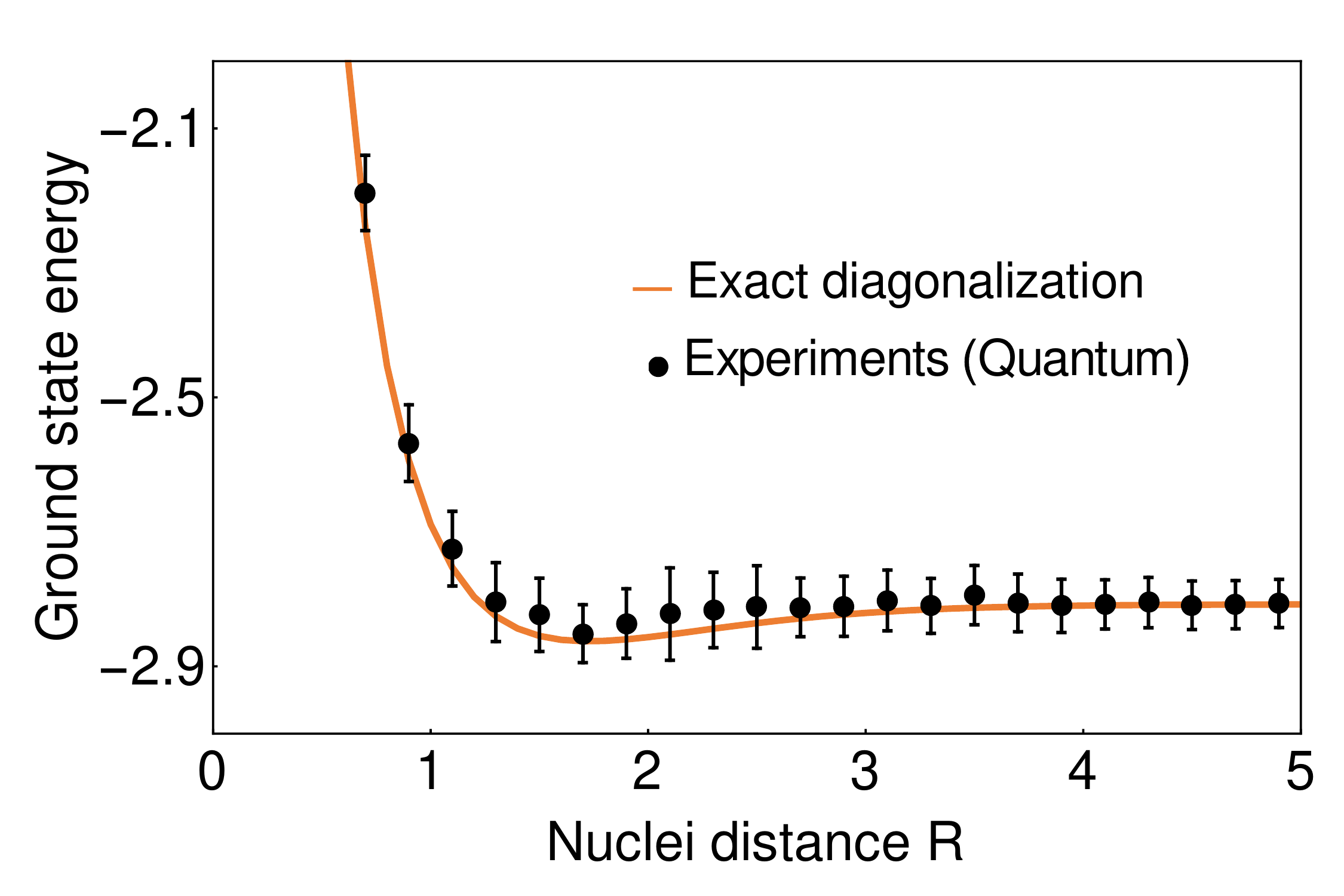}
\caption{\label{fig:Fig3}The ground state energy of HeH$^+$ depending on the inter-nucleus distance $R$. The error-bars of the experimental data mainly come from the quantum projection noise of 1000 repetitions for each term of the Hamiltonian (\ref{eq3}).}
\end{figure}

Fig.~\ref{fig:Fig2} shows an instance of the energy optimization process, when the nuclei separation of HeH$^+$ is fixed to be $R=1.7~ a.u.$. Note that throughout the paper, the atomic unit ($a.u.$) is used. The algorithm is capable of finding the minimum energy and state in around hundred iterations with the full six-parameter simulations as shown in Fig.~\ref{fig:Fig2}(a). About twice less iterations shown in Fig.~\ref{fig:Fig2}(b) can be achieved for an ansatz simplified to contain two parameters (see SM. F: Reduction of Parameters). Since both cases provide equivalent results, we focus on the two-parameter ansatz in the following discussion. Figs.~\ref{fig:Fig2}(c)(d) show the typical search of minimum energy by the classical Nelder-Mead algorithm with two parameters. 

Fig.~\ref{fig:Fig3} shows the energy curve of the ground state of  HeH$^+$ depending on the nuclear distance $R$, where each point is obtained by the procedure of Fig.~\ref{fig:Fig2}. The experimental data are in agreement with the energy (orange line) calculated by the exact diagonalization of the full matrix of Hamiltonian~(\ref{eq3}) within the error bars. From the energy curve, the equilibrium distance between the nuclei is located at $R=1.73 a.u.$ with the corresponding energy of $E=-2.86\pm0.05~a.u.$.

\begin{figure}[ht]
\includegraphics[width=0.5\textwidth]{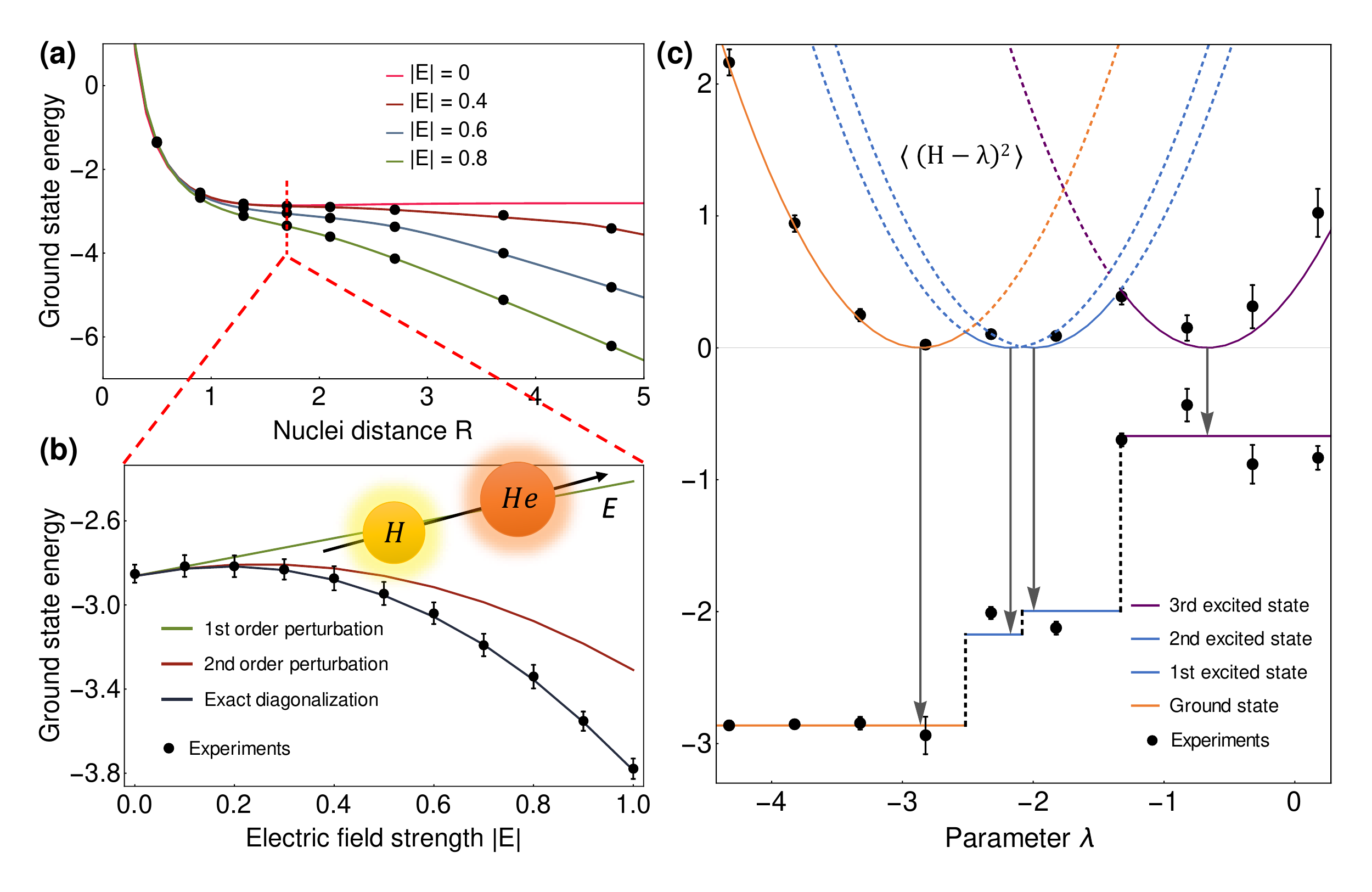}
\caption{\label{fig:Fig4} Applications of the UCC simulation. (a) The ground state energy of HeH$^+$ subject to a static electric field along the nuclei axis for different strengths. (b) The comparison between the UCC quantum simulation and the perturbation theory at given inter-nucleus distance $R=1.7 a.u.$. (c) The search process of the energies of excited states of $H$ by finding the ground-state energy of the Hamiltonian $(H-\lambda)^2$ by scanning the values of $\lambda$. When $\lambda$ is the energy of an excited state, the experimental minimum value of $\avg{(H-\lambda)^2}$ tends to be zero. We also can calculate the excited-state energy from other non-zero values of $\avg{(H-\lambda)^2}$. If $\lambda$ is on the left (right) side of the excited energy, the positive (negative) solution of $E_{\rm meas} = \langle (H-\lambda)^2 \rangle$  provides the excited energy.}
\end{figure}

Furthermore, the same procedure can be used to study the non-perturbative behaviors of the HeH$^+$ molecular ion under strong electric field with the new target Hamiltonian including the effect of the electric field as $\mathbf{E} \cdot ({{\mathbf{r}}_{{\text{1}}}} + {{\mathbf{r}}_{{\text{2}}}} ) - \mathbf{E} \cdot ({{2\mathbf{R}}_{{\text{He}}}} + {{\mathbf{R}}_{{\text{H}}}} )$ (see SM. F: The Electric Field Effect on HeH$^+$ ). Fig.~\ref{fig:Fig4}(a) shows the phenomenon of chemical-bond softening of HeH$^+$ (at $R=1.7 ~a.u.$) as the strength of the electric field increases, which eventually leads to a dissociation of the molecular ions \cite{Sheehy1996}. We compare our non-perturbative results with those obtained through the first-order and second-order perturbation theories shown in Fig.~\ref{fig:Fig4}(b).

Finally, we also study the excited states of $H$ by changing the target Hamiltonian to $(H-\lambda)^2$, where $\lambda$ is a parameter close to the energy of an excited state, which turns the excited state of $H$ into the ground state of $(H-\lambda)^2$ (see SM. G:The Computation of Excited States Energy). In the experiment, we uniformly scan the values of $\lambda$ and apply the same UCC procedure to find the minimum energy in a given $\lambda$. As shown in Fig.~\ref{fig:Fig4}(c), we observe that the required precision for the computation of excited states should be much higher than the separation of the energies. In the current limited system, we obtain the energy of the highest excitation that has relatively large energy gap to other states but the rest of them are not well resolved.

Our current realization is capable of simulating any molecule up to four electronic levels with a single ion. In general, a molecule of $N$ electrons system in $M$ molecular orbitals ($M\geq N$) can be implemented with $M$ qubits system or $M/2$ qudits, four-level systems shown in our realization, through the Jordan-Wigner transformation and four-level mapping. For the UCC implementation with $M$ qubit system, it requires the simulation of time-evolution of $M$-body interaction, which is equivalent to the nonlocal product of $M$ Pauli operators. The simulation of such $M$-body interaction, which is the most challenging operation in the UCC protocol, can be performed by applying $2M$ times of CNOT-gate or 2 times of the multi-particle M{\o}lmer-S{\o}rensen gates \cite{Yung2014,Casanova2012,Lanyon2011}. The measurement of $M$-qubit Hamiltonian with the $O(M^{4})$ terms has been already well established in the trapped ion system. For the $M/2 $-qudit Hamiltonian, we can simply use the same measurement scheme used in our experimental demonstration. The UCC scheme for the trapped ions can be applied to other physical platforms \cite{Houck2012,Aspuru-Guzik2012a,Georgescu2014}.    

We emphasize the computational complexity of the quantum implementation of the UCC method scales polynomially with the number of orbitals $M$. Including the maximum excitation up to $k$, each cluster operator contains $k$ creation operators and $k$ annihilation operators. For a total of $M$ orbital modes, therefore, we have a total of $O(M^{(2k)})$ terms. After the Jordan-Wigner transformation, the fermionic operators are mapped into spin operators, which requires $O(M)$ operations. The total number of scaling as the number of molecular orbits $M$ is $O(M^{(2k+ 1)})$. Moreover, the time evolutions and the measurements in our UCC implementation allow parallel computation \cite{Yung2014,Peruzzo2013}, which boosts the performance. Our experimental realization of UCC method opens a new dimension of quantum simulation and offers a solution for the classical coupled-cluster methods. We note that some of other current developments and understandings in the coupled-cluster schemes could be adapted in quantum UCC scheme. Moreover, our UCC scheme could be applied to other large eigenvalue problems in network search algorithm and condensed matter physics.

\textbf{Notes}
This paper is a revised version of [arXiv:1506.00443] by the same authors. While the work is under consideration for publication, the authors noted that a paper reporting Scalable Quantum Simulation of Molecular Energies in Phys. Rev. X 6, 031007 (2016).

\begin{acknowledgments}
We thank Cheol Ho Choi, Kyoung Koo Baeck, Ryan Babbush and Jarrod McClean for the helpful discussion. This work was supported by the National Key Research and Development Program of China under Grants No. 2016YFA0301900 (No. 2016YFA0301901), the National Natural Science Foundation of China 11374178, 11405093, 11574002 and 11504197. M.-H.Y. and K.K. acknowledge the recruitment program of global youth experts of China.
\end{acknowledgments}

%

\end{document}